\documentclass[12pt,a4paper]{article}
\usepackage{jheppub}
\usepackage[T1]{fontenc} 
\usepackage{epsfig,multirow,subfigure} 
\usepackage{amscd}
\usepackage[matrix,arrow,curve]{xy}
\usepackage{verbatim}
\usepackage{latexsym}
\usepackage{amsfonts,amsthm,amsmath,mathtools}

\newcommand{\sgn}{{\rm sgn}}
\newcommand{\tr}{{\rm tr}}
\newcommand{\md}{{\rm mod}}

\title{\centering
Chern-Simons and RG Flows: \\Contact with Dualities
}
\author[a]{Antonio Amariti,}
\author[b,c]{Claudius Klare,}

\affiliation[a]{Laboratoire de Physique Th\'eorique de l'\'Ecole Normale Sup\'erieure \\
24 Rue Lhomond, Paris 75005, France}
\affiliation[b]{Institut de Physique Th\'eorique, CEA/Saclay \\
CNRS URA 2306, 
F-91191 Gif-sur-Yvette, France} 
\affiliation[c]{Institut des Hautes \'Etudes Scientifiques\\
35 Route de Chartres, F-91440 Bures-sur-Yvette, France}
               
\emailAdd{amariti@lpt.ens.fr}
\emailAdd{claudius.klare@cea.fr}

\abstract{Contact terms in two point functions of global symmetry
  currents have recently been proposed as a check of Seiberg-like
  duality in three dimensional supersymmetric field theories.  In this
  paper we compute the contact terms for various $\mathcal{N}=2$ dual
  pairs in flat space.  We show that the results of this computation
  agree with the ones obtained from localization.  We study dual pairs
  of gauge theories with (anti-)fundamental matter fields, and some
  special examples of dual pairs with adjoint and antisymmetric matter
  fields.  We also propose a duality between unitary and
  symplectic gauge theories.  }

\begin{document}

\maketitle
%
%
%
%
%
%
%
%
%
%
%
%
%
\section{Introduction}
\label{Sec:intro}

Describing the strong coupling regime is one of the main complications
when studying the low energy behaviour of quantum field theories.
Because the perturbative approach breaks down some non-perturbative analysis is necessary.  

For instance, one can sometimes find a dual picture in terms of new
degrees of freedom, describing the same physics.  A
popular example is the electric-magnetic duality introduced by Seiberg in \cite{Seiberg:1994pq} which provides a dual description of
$\mathcal{N}=1$ SQCD.   

A closely related setup is the $\mathcal{N}=2 $ three dimensional case,
that can be obtained by dimensional reduction from $\mathcal{N}=1$ in four dimensions.
It is natural to expect an extension of Seiberg-duality in this setup.
However, the three dimensional dynamics is quite different and finding Seiberg-like dualities was not straightforward. 

A first possibility was discussed by Aharony in
\cite{oai:arXiv.org:hep-th/9703215}. This duality maps two three dimensional SQCD like
theories; the dual magnetic theory includes extra degrees of freedom with respect to the four dimensional cousin. 
After that the subject has not been pursued for almost ten years, mainly for lack of non-perturbative techniques in three dimensional theories.
In the last years there have been important results in this direction.
First, the discovery of the AdS$_4$/CFT$_3$ duality \cite{Aharony:2008ug}  provided a holographic description of the strongly coupled dynamics.
Then, localization gave access to quantum exact results \cite{Pestun:2007rz}, which extend to the non-perturbative regime.

Another duality, between three dimensional SQCD like theories with Chern-Simons (CS) interaction,
was discovered by Giveon and Kutasov in \cite{oai:arXiv.org:0808.0360}.
This duality was derived from the brane transition of \cite{Hanany:1996ie} 
and can be generalized to quiver gauge theories \cite{Aharony:2008gk,Amariti:2009rb}.
It was then shown \cite{Willett:2011gp} that Giveon-Kutasov duality
can also be obtained by an RG flow from Aharony duality.\footnote{Recently
the existence of ``reverse'' flows has been discovered too \cite{Intriligator:2013lca,Khan:2013bba,Amariti:2013qea}.
We will turn back to this point in section \ref{sec:contact-unitary} and in appendix \ref{sec:APPFLOWS}.}
Similar flows have been applied in \cite{Benini:2011mf} to generate new classes of dualities.
Furthermore other generalizations of Seiberg-like dualities, with a richer matter content, have been studied 
\cite{oai:arXiv.org:0808.2771,oai:arXiv.org:0903.0435,Jafferis:2011ns,Morita:2011cs,oai:arXiv.org:1110.2547,Kim:2013cma}.

Standard techniques in four dimensions, as 't Hooft anomaly matching 
and $a$-maximization \cite{Intriligator:2003jj}, do not exist in three dimensions.
However, a powerful technology for analyzing the IR dynamics of three dimensional supersymmetric field theories
has recently been provided by localization.
The partition function on some compact manifolds as $S^2 \times_w S^1$ and $S^3$
has been reduced to a matrix integral \cite{Kapustin:2009kz,Kim:2009wb,Jafferis:2010un,Hama:2010av}.
In this paper we will be mostly interested in the partition function on the squashed three-sphere $S^3_b$ \cite{oai:arXiv.org:1102.4716},
involving integrals of ``hyperbolic hypergeometric functions''.
Recently non-trivial identities among these integrals have been derived in the mathematical literature \cite{VdB};
quite remarkably in \cite{Willett:2011gp,Benini:2011mf} it has been shown that some of these identities relate 
the partition functions of the dual pairs discussed above.
This provides a highly non-trivial check of Seiberg-like dualities in three dimensions.

However these identities contain an extra phase, which did not have an immediate physical interpretation. 
Interestingly, it was found in \cite{Benini:2011mf} that this phase corresponds to CS terms for global symmetries, 
generated when flowing from Aharony duality. 
A more intrinsic interpretation of the CS terms was then given in \cite{Closset:2012vg,oai:arXiv.org:1206.5218}.
It was observed that they are related to contact terms in the two point functions of global symmetry currents.
This represents a new check of dualities, 
in a very broad sense reminiscent of 't Hooft anomaly matching in four dimensions.  

In this paper we apply this check to various dual pairs and match the
result with the one obtained from localization.
We first analyze the dualities studied in \cite{oai:arXiv.org:hep-th/9703215,oai:arXiv.org:0808.0360,Benini:2011mf},
then we study certain generalizations of \cite{Jafferis:2011ns,oai:arXiv.org:1110.2547}.
Moreover we propose some new dualities between
unitary and symplectic gauge theories and test them both with
localization and by computing the contact terms.

The paper is organized as follows. In section \ref{Sec:general} we
review relevant aspects of $\mathcal{N}=2$ theories in three
dimensions. In section \ref{sec:contact} we introduce the contact
terms of two point functions and their relation with dualities.
We study them in a simple example.
In section \ref{sec:contact-unitary} we discuss gauge theories with
fundamental matter. In section \ref{newtoto} we study gauge theories with
adjoint matter, that are dual to theories without gauge groups.
In section \ref{Sec:newduality} we argue for a new duality between an unitary and
a symplectic theory. 
In appendix \ref{sec:Part-func} we review some aspects of
the partition function on the squashed three sphere, 
summarize the integral identities that we used in the paper and discuss the
relation between the partition function and the contact
terms. 
In appendix \ref{sec:APPFLOWS} we discuss some RG flow between dual pairs,
generalizing the reverse flows of \cite{Intriligator:2013lca,Khan:2013bba,Amariti:2013qea}.
%
%
%
%
%
%
%
%
%
%
%
%
%
\section{Aspects of $\mathcal{N}=2$ three dimensional theories}
\label{Sec:general}

In this section we review some general aspects of $\mathcal{N}=2$
three dimensional gauge theories that we will use in the rest of the
paper.

These theories have four supercharges as $\mathcal{N}=1$ theories in
four dimensions and the supersymmetric representations are closely related. 
However there are some differences. 
There is an additional real scalar $\sigma$ in the vector multiplet.
A generic non zero vev for this scalar breaks the gauge group $G$ to $U(1)^r$, where $r$ is the rank of $G$.

Another extra feature in three dimensions is that one can dualize the photon into a scalar,
$F=*d\phi$.
This allows to trade abelian vector multiplets by chiral multiplets with lowest component $e^{\Phi}=e^{i \phi + \sigma}$.
The Coulomb branch $U(1)^r$ can hence be parameterized by $r$ chiral multiplets $e^\Phi_i$.  
In the UV they correspond to monopole operators.
Furthermore, there are $r$ symmetries $U(1)_{J_i}$ related to the topological currents $J_i = *F_i$, conserved by the Bianchi identity.
These symmetries shift the corresponding dual photons and hence the monopole operators are charged under them.
At quantum level, most of the classical Coulomb branch is lifted by a superpotential generated by instantons \cite{Affleck:1982as}.
However, for theories with matter fields, typically some directions of the Coulomb branch remain unlifted
and one topological $U(1)_J$ survives.
Another global symmetry that plays an important role in our analysis is the
$SO(2)_r \simeq U(1)_r$ $R$-symmetry that rotates the supercharges.
Moreover in three dimensions there are no anomalies and \emph{e.g.\ }axial symmetries are allowed in the quantum theory.

Another important characteristic of three dimensional theories is the
appearance of the topological CS action for the vector multiplet
\begin{equation}
  \label{eq:CS-action}
S_{CS} = \frac{k}{4\pi}  \int \tr (
A \wedge dA + \frac{2}{3}A^3 - \lambda \widetilde \lambda + 2 \sigma D )
\end{equation}
Invariance under global gauge transformations constrains the CS level $k$ to be integer \cite{Deser:1981wh}.\footnote{See 
also \cite{oai:arXiv.org:1206.5218} for a recent discussion.}

We can turn on real masses for the chiral fields through their coupling to the global symmetries.
By weakly gauging the symmetries and assigning a vev to the scalars $\sigma_i$ in the $i$-th background multiplet
we generate the mass term
$ \int d^4 \theta X^{\dagger} e^{\mu \theta \overline \theta} X $.
Here
$\mu=\sum q_i m_i$, where $m_i = \langle \sigma_i \rangle $ and $q_i$ is the charge of $X$ under $U(1)_i$.
Also the topological $U(1)_J$ can contribute to the real mass.
In this case the vev of the background scalar appears as an FI term for the gauge multiplet.

Integrating out fermions with a real mass shifts the CS level of the global and local symmetries.
These shifts are generated at one loop by
\begin{equation}
\label{1loopdermions}
k^{\text{eff}}_{ij} = k_{ij} +
\frac{1}{2} \sum q^{\psi}_i q^{\psi}_j sgn(\mu_\psi)
\end{equation}
where $k_{ij}$ is the bare CS level for the two symmetries $U(1)_i$ and $U(1)_j$ and
the sum is over the massive fermions $\psi$ with real masses $\mu_\psi$.
Since $k_{\text{eff}}$ has to be integer, it follows that the bare CS level is quantized in units of  $\frac12 \sum q^\psi_i q^\psi_j \; \md \; 1$
and possibly parity is broken \cite{Niemi:1983rq,Redlich:1983dv}.
Note that for non-abelian symmetries we have
\begin{equation} 
\label{eq:k-non-abelian}
k^{\text{eff}} = k+ \frac12 \sum \sgn(\mu_\psi) T_2(R_\psi)
\end{equation} 
where $T_2(R_\psi)$ is the quadratic index of the representation $R_\psi$ of the fermion $\psi$.

\section{Contact terms and duality}
\label{sec:contact} 

In this section we outline the basic logic of this paper.
We first discuss the contact terms that appear in two point functions of global symmetry currents
and their usefulness for checking dualities. 
We review some of their properties discussed in \cite{oai:arXiv.org:1206.5218} 
and summarize the relation with duality and localization.
Finally we elucidate the strategy by discussing an instructive example in subsection \ref{sec:simple}.

Consider two global $U(1)$ currents $j_{\mu}^i$ and $j_{\nu}^j$ in a three dimensional QFT. The
correlator of the two point function is
\begin{equation}
\langle j_{\mu}^i(x) \ j_{\nu}^j(0) \rangle = 
\left(\delta_{\mu \nu}\partial^2-\partial_\mu \partial_\nu\right)
\frac{\tau_{ij}}{32 \pi^2 x^2}
+
\frac{i k_{ij}}{2 \pi} \epsilon_{\mu \nu \lambda} \partial^{\lambda}\delta^{(3)}(x)
\end{equation}
The first term is the usual correlator of two point functions, governing the physics at separated points.
The second term contains a delta function and hence is a contact term, describing the behavior at coincident points.
In the action it is associated to a CS term $k_{ij}$ for the background gauge fields of the global symmetries $j_\mu^i$ and $j_{\nu}^j$.

The study of contact terms in three dimensions is interesting for
various reasons \cite{Closset:2012vg,oai:arXiv.org:1206.5218}.  The fractional part of $k$ is an
unambiguous observable of the field theory and a
non-vanishing fractional part of some counterterms lead to a new
anomaly in superconformal field theories. Finally, global CS terms are also useful in the
analysis of Seiberg-like dualities and this is the focus in the following.

\subsubsection*{Localization} 
The partition function of an $\mathcal{N}=2$ three dimensional theory
on the squashed three sphere $S^3_b$ can be exactly computed at one loop by
localization. Here $b$ denotes the squashing parameter \cite{oai:arXiv.org:1102.4716}.
The final result is an expression for the path integral
in terms of a matrix model involving some integrals of ``hyperbolic
hypergeometric functions'' as reviewed in the appendix
\ref{sec:Part-func}.  In the recent mathematical literature integral
identities for these functions have been studied \cite{VdB}. 
It was observed \cite{Willett:2011gp,Benini:2011mf} that these identities relate the
partition functions of pairs of dual field theories.

There is an extra phase in these relations, that is crucial for the forthcoming.
It arises in localization from the global CS couplings associated with the contact terms.
The supersymmetric saddle point for the background multiplets is $D_I = i \sigma_I$ and a vev $\langle \sigma_I \rangle = m_I$
of the background real scalars, with Lagrangian \eqref{eq:CS-action}, gives rise to a phase in the partition function.
Note that for preserving supersymmetry on the three sphere we need to switch on an imaginary background for the $R$-symmetry
\cite{Festuccia:2011ws}, which makes this phase actually complex.

\subsubsection*{Dualities}
In \cite{Benini:2011mf} the authors derived this phase in a physical picture by constructing various dualities 
through flowing in from Aharony duality.
Integrating out matter generates the CS terms for the global symmetries and hence determines the phase.
Matching this result with the mathematical identities is an additional check.

The alternative approach of \cite{oai:arXiv.org:1206.5218} is related to the contact terms.
Indeed the contact terms of the two point functions are related to CS terms for the global symmetries that appear in the action.
Integrating out all matter brings the theory to a purely topological one,
for which the contact terms have been computed \cite{oai:arXiv.org:1206.5218}.
By adding the $1$-loop CS couplings, which are generated along the flow, as counterterms to the UV duality
one reconstructs the contact terms of the topological theory. 
\\
\\

At this point of the discussion we can summarize our strategy. 
For a given dual pair we integrate out the matter fields, generating global CS terms via $\eqref{1loopdermions}$.
In this way we flow to a duality between two topological theories.
Combining the $1$-loop CS terms with the contact terms of the topological theory we determine the relative contact terms
of the original duality.
Finally we check that these match with the complex phase that appears in the partition function on the three sphere.

\subsection{A warm-up example}
\label{sec:simple}

In this subsection we discuss a simple but instructive example, containing the typical technical aspects of the computations in this paper.
It is a subcase of the ``$(p,0)$-duality'' studied in \cite{Benini:2011mf} and analyzed in section \ref{sec:contact-unitary}.
Physically, this duality is needed when generating monopoles along an RG-flow; indeed we will use the results obtained here in
the third part of section \ref{sec:contact-unitary}.

On the electric side we have a chiral $U(k)_{k/2}$ gauge theory
with $k$ fundamental and no anti-fundamental flavors.
This theory is dual to a gauge singlet \cite{Dimofte:2011ju,Benini:2011mf,Beem:2012mb,Intriligator:2013lca,Amariti:2013qea}, 
corresponding to the monopole of the electric theory with magnetic flux $(1,0,\dots,0)$ in the Cartan of the gauge group.
There are three global $U(1)$ symmetries: the axial $U(1)_A$, the
$R$-symmetry $U(1)_R$ and the topological $U(1)_J$.

We start by computing the contact terms of the electric theory.  Each
chiral field has charge $1$ under the axial symmetry and $R$ charge
$\Delta$. We consider an effective theory at scales much lower than
the real masses of the quarks.  
As \emph{modus operandi} in this note we choose the sign of the masses such that the effective CS coupling does not hit zero.
Hence we consider positive real masses for the quarks when integrating them out.
This generates via $\eqref{1loopdermions}$ at $1$-loop the global CS terms
\begin{eqnarray}
k_{rr}^e = \frac12 k^2 (\Delta\!-\!1)^2 \quad \,
k_{rA}^e = \frac 12 k^2 (\Delta\!-\!1)\quad \,
k_{AA}^e = \frac12 k^2\quad  \,
k_{AG}^e = k\quad \,
k_{rG}^e = k (\Delta\!-\!1)\nonumber \\
\end{eqnarray}
Note that the fermion of the chiral multiplet has $R$ charge $\Delta-1$,
in total it has $k$ colour and $k$ flavour components.
After this step we are left with a pure CS gauge theory $U(k)_k$.
Due to the chiral matter content, we generated also the mixed global-gauge
CS couplings $k_{AG}$ and $k_{rG}$.
In the Lagrangian, they give rise to FI terms of the gauge group $G$
\begin{equation}
  \delta \mathcal{L} = (k_{AG} m_A + k_{rG} \omega) \tr D
\end{equation}
Here $\omega$ is related to the background vev for the $R$ symmetry. 
On a curved space it has to be properly chosen in order to preserve supersymmetry,
after embedding the $R$ symmetry in a background gravity multiplet \cite{Festuccia:2011ws}.
For the case of the squashed sphere $S^3_b$ we have $\omega = \frac i2 (b + \frac1b)$ \cite{oai:arXiv.org:1102.4716}.
Recall that the real mass parameter for the $U(1)_J$ symmetry also appears
in the Lagrangian as an FI term $-\frac12 \lambda \, \tr D$.  We can
combine the $U(1)_J$ symmetry with the axial and the $R$-symmetry to
absorb the mixed terms $k_{AG}$ and $k_{rG}$.  In other words, in the
IR we have an effective topological symmetry
\begin{equation} 
  U(1)_{\widetilde J} = U(1)_J -  k_{AG} U(1)_A -  k_{rG} U(1)_r
\end{equation}
with a ``real'' mass parameter that is related to the original parameters as
$\tilde \lambda = \lambda -2k_{AG} m_A -2 k_{rG} \omega$. 
Note that the real mass for $U(1)_J$ is $\frac12 \lambda$.

Finally, the pure CS theory gives two extra contributions to
$k_{rr}^e$ and to $k_{\widetilde J \widetilde J}^e$ as explained in
\cite{oai:arXiv.org:1206.5218}.  They are
\begin{align}
  k_{rr}^{e, \lambda}=-\frac12 k^2  &&
k_{\widetilde J \widetilde J}^e=-1
\end{align}
where $k_{rr}^{e, \lambda}$ comes from integrating out the gaugini
$\lambda$ with $R$ charge $1$ and topological mass proportional to $-k$. 
Also, $k_{\widetilde J \widetilde J}^e$ is the CS coupling of the effective theory for the background vector describing the
$U(1)_{\widetilde J}$ symmetry \cite{oai:arXiv.org:1206.5218}.

On the magnetic side we have a singlet $T$ with $R$ charge
$\Delta_T=1-\frac{k}{2}(1+\Delta)$, axial charge $-k/2$ and
topological charge $1$.  
It is dual to the electric monopole and its $U(1)$ charges can be computed by a $1$-loop computation from the fermion spectrum 
\cite{Benna:2009xd,Jafferis:2009th,Benini:2011cma}.
By integrating out $T$ we get
\begin{equation}
\begin{array}{cc}
\begin{array}{ccl}
k_{rr}^m &=& -\frac{1}{2} (\Delta_T-1)^2 \\
k_{rA}^m &=& \frac{1}{4} k (\Delta_T-1) \\
k_{AA}^m &=& -\frac{1}{8}k^2  \\
\end{array}
& \hspace{3cm}
\begin{array}{ccl}
k_{rJ}^m &=& -\frac{1}{2}(\Delta_T-1) \\
k_{AJ}^m &=& \frac{1}{4} k  \\
k_{JJ}^m &=& -\frac{1}{2} \\ 
\end{array}
\end{array}
\end{equation}
Eventually, by combining the electric and the magnetic results, in
terms of the original symmetries we have 
the following counterterms to add to the magnetic theory
\begin{equation}
  \begin{aligned}
\label{eq:contact-Uk-k2}
\Delta k_{rr} &= -\frac{1}{8} (1+2 \Delta-3 \Delta^2) k^2 
&\Delta k_{rJ} &= 
-\frac{1}{4} k(1-3 \Delta  )\\
\Delta k_{AA} &= 
\frac{3}{8} k^2 
&\Delta k_{AJ} &= 
\frac{3}{4} k\\
\Delta k_{rA} &= 
-\frac{1}{8}(1-3 \Delta) k^2
&\Delta k_{JJ} &= 
-\frac{1}{2}
\end{aligned}
\end{equation}
where $\Delta k_{rr} = k_{rr}^e +k_{rr}^{e, \lambda}-k_{rr}^m + k_{\widetilde J
\widetilde J} (k_{rG})^2 $ etc.

We can check the validity of this computation by comparing with the results from localization.
Indeed via $\eqref{eq:phase}$ the global CS couplings reproduce the phase
in the integral identity that relates the partition functions of the duality \cite{Benini:2011mf}.

\section{Dualities with fundamental matter}
\label{sec:contact-unitary}
In this section we compute the CS counterterms for various dualities 
with unitary gauge group and matter fields transforming in the fundamental representation.
In \cite{oai:arXiv.org:1206.5218} this study has been initiated for Giveon-Kutasov duality, 
here we will extend this check to other dualities.
\subsubsection*{Aharony duality}
This generalization of four dimensional Seiberg duality was introduced in \cite{oai:arXiv.org:hep-th/9703215}.
\begin{itemize}
\item 
  The electric phase is a $U(N_c)_0$ gauge theory without CS term, 
  $F$ pairs of fundamentals and antifundamentals $Q$, $\tilde Q$ and vanishing
  superpotential.
\item
  The magnetic dual is a $U(\widetilde N_c)_0$ gauge theory with $\widetilde N_c = F-N_c$.
  There are $F$ dual (anti-)fundamentals $q$ and $\tilde q$.  As in four
  dimensional Seiberg duality there is also a meson $M=Q \tilde Q$.
  However, in three dimensions, there are two additional gauge
  singlets $T$ and $\widetilde T$, corresponding to the monopole and the
  antimonopole of the electric theory, with magnetic flux $(\pm 1,
  0,\dots,0)$.  These singlets couple to the magnetic monopoles $t,
  \tilde t$. The superpotential is
\begin{equation}
  W_{dual} = M q \tilde q + t T + \tilde t \widetilde T 
\end{equation}
\end{itemize}
The charges of the fields under the global symmetries are summarized in table $\eqref{eq:charges-table-fundam}$.
Here we briefly report that the relative contact terms in the duality identically vanish.
Integrating out the matter fields we find
\begin{equation}
\begin{aligned}
  &k_{RR}^e = N_c F (\Delta-1)^2 \\
  &k_{RR}^m = - \widetilde N_c F \Delta^2+ \frac12 F^2 (2\Delta -1)^2 - (F(1-\Delta)-N_c)^2 \equiv k_{RR}^e -\frac12 (N_c^2+ \widetilde N_c^2) \\
 & k_{RA}^e = N_c F (\Delta-1) \\
 & k_{RA}^m = - \widetilde N_c F \Delta+ F^2 (2\Delta -1) + F (F(1-\Delta)-N_c) \equiv k_{RA}^e \\
 & k_{AA}^e = N_c F \hspace{3.7cm}
  k_{AA}^m = - \widetilde N_c F + 2 F^2 - F^2  \equiv k_{AA}^e\\
 & k_{JJ}^e = 0 \hspace{4.4cm}
  k_{JJ}^m = -1 \\
 & k_{SU(F)_L}^e = k_{SU(F)_R}^e= \frac{N_c}{2} \hspace{1.5cm}  
  k_{SU(F)_L}^m = k_{SU(F)_R}^m= -\frac12 (\widetilde N_c + F )
\end{aligned}
\end{equation}
We see that the relative global CS terms which are generated along the flow directly reproduce the 
contact terms of the topological theory \cite{oai:arXiv.org:1206.5218} and no counterterms are required.

\subsubsection*{$(p,0)$ and $(p,q)$-duality}
In this section we consider the $(p,0)$ and the $(p,q)$ Seiberg-like dualities 
with chiral field content introduced in \cite{Benini:2011mf}.
In that paper the contact terms for the global symmetries have been determined while
deriving these dualities through a flow from Aharony duality, giving masses to a set of chiral matter fields.
Here we reproduce this result by calculating the contact terms as described above.
Let us give some details on the $(p,0)$-duality, the $(p,q)$-case is very similar.
\begin{itemize}
\item 
  The electric theory has gauge group $U(N_c)_{k}$ with
  $k=-\frac{F_R-F_L}{2}$ and $F_R>F_L$. There are $F_L$ fundamental
  and $F_R$ antifundamental fields.
\item The magnetic theory is an
  $U(F_R-N_c)_{-k}$ gauge theory with $F_R$ fundamental and $F_L$
  antifundamentals. The singlets are the meson $M=Q \tilde Q$  and a
  monopole $T$.  There is a superpotential
\begin{equation}
  W_{dual} = M q \tilde q + t T 
\end{equation}
\end{itemize}
The charges under the global symmetries are found in table $\eqref{eq:charges-table-fundam}$.
There is a small extra complication with respect to Aharony duality,
since integrating out chiral matter generates mixed global-gauge CS terms 
\begin{equation}
  k_{RG}^e = k_{AG}^e (\Delta-1) = -\frac12 (F_R-F_L)
\end{equation}
and similarly, but with opposite sign, in the magnetic sector.
As discussed in section \ref{sec:simple}, these mixed CS couplings will shift the IR topological symmetry as
$U(1)_{\widetilde J} = U(1)_J +\frac12 (F_L + F_R) (U(1)_A + (\Delta-1) U(1)_r)$.
By integrating out the fermions we find
\begin{equation}
\begin{aligned}
  k_{RR}^e &= \frac12 (- N_c (F_L + F_R) (\Delta-1)^2 )\\
  k_{RR}^m &= 
  \frac12 (\widetilde N_c (F_L+F_R) \Delta^2+ F_L F_R (2\Delta -1)^2 + (\Delta_T-1)^2 )\\
  k_{RA}^e &= -\frac12 N_c (F_L + F_R) (\Delta-1) \\
  k_{RA}^m &= \frac12 ( \widetilde N_c (F_L+F_R) \Delta- F_L F_R (2\Delta -1) - \frac12 (F_L+F_R) (\Delta_T-1)) \\
  k_{AA}^e &= -\frac12 N_c (F_L + F_R) \\
  k_{AA}^m &= \frac12 ( \widetilde N_c (F_L + F_R) - 4F_L F_R  - \frac14 (F_L + F_R)^2) \\
  k_{JJ}^m &= \frac12  \quad \quad \quad
  k_{AJ}^m = \frac14 (F_L + F_R) \quad \quad \quad
  k_{rJ}^m = -\frac12 (\Delta_T -1)
\end{aligned}
\end{equation}
and a pure gauge theory on both sides of the duality.
We can integrate out also the gaugini 
and from the effective Lagrangian for the shifted topological $U(1)_{\widetilde J}$ we get \cite{oai:arXiv.org:1206.5218}
\begin{align}
  &k_{\widetilde J \widetilde J}^e = -\frac{N_c}{k - \frac12 (F_L + F_R)}
  & k_{\widetilde J \widetilde J}^m = -\frac{\widetilde N_c}{-k + \frac12 (F_L +F_R)} 
\end{align}
As counterterms on the magnetic side, the $1$-loop terms reproduce via $\eqref{eq:phase}$ precisely the complex phase 
in the partition functions of the $(p,0)$-duality in \cite{Benini:2011mf}.
Note that the real mass parameter for $U(1)_J$ is $\frac12 \lambda$ while the one for $U(1)_{\widetilde J}$ is 
$\frac12 (\lambda + (F_L + F_R) (m_A + (\Delta-1) \omega))$.

Given the contact terms of Aharony duality above, our check is complementary to the analysis in \cite{Benini:2011mf}.
We have also checked that a very similar analysis reproduces the contact terms of $(p,q)$-duality, 
here we skip the details.

\subsubsection*{$(p,q)^*$ -duality}
The $(p,q)^*$ duality introduced in \cite{Benini:2011mf} requires some more comments. 
\begin{itemize}
\item 
  The electric theory is a $U(N_c)_{k}$ gauge theory with
  $k=f-\frac12 (F_L+F_R)$ and $F_R>f>F_L$. There are $F_L$
  fundamentals $Q$ and $F_R$ antifundamentals $\widetilde Q$.
\item The magnetic theory is a $U(F_R-N_c)_{-k}$ gauge theory 
  with $F_R$ fundamentals $\tilde q$, $F_L$ antifundamentals $q$ and a meson $M=Q \tilde Q$. 
  There is a superpotential
\begin{equation}
  W_{dual} = M q \tilde q 
\end{equation}
\end{itemize}
The charges under the global symmetries are displayed in table $\eqref{eq:charges-table-fundam}$.
Along the RG flow to the topological theory we run into the difficulty that at some point the effective CS level vanishes. 
In such a situation extra fields appear and the gauge group is broken to some smaller rank \cite{Intriligator:2013lca},
see also \cite{Khan:2013bba,Amariti:2013qea}.
We hence reduce the dual pair to the vector like Aharony duality,
for which we have computed the contact terms above.
The flow can be achieved by giving a large negative mass to $f-F_L$ and a large positive mass to $F_R-f$ of the electric antiquarks.
Integrating out the heavy fermions, the effective CS coupling vanishes.
We are left with $F_L$ pairs of (anti-)fundamentals $Q,\widetilde Q$.

In the magnetic dual also $F_R-F_L$ of the adjoint scalars in $U(F_R-N_c)$ get a large vev,
similar to the cases studied in \cite{Intriligator:2013lca,Khan:2013bba,Amariti:2013qea},
$f-F_L$ with a negative and $F_R-f$ with a positive sign.
This breaks the dual gauge group to $U(F_L-N_c) \times U(f-F_L) \times U(F_R-f)$.
We choose the large vevs of the gauge scalars such that 
$F_L$ quark-antiquark pairs $(q,\tilde q)$ of the $U(F_L-N_c)$ sector,
$f-F_L$ antiquarks in the $U(f-F_L)$ sector and $F_R-f$ antiquarks in the $U(F_R-f)$ sector remain massless.
At the same time, the spectrum of heavy quarks is 
\begin{equation} 
  \label{eq:pqstar-masses-q}
\begin{array}{ccc|cccc|c}
                     U_{F_L-N_c}  & U_{f-F_L}   &   U_{F_R-f} &   SU_{F_R-f}^L &   SU_{f-F_L}^L &   SU_{F_L}^L &   SU_{F_L}^R & sgn(\mu)\\
\hline
\Box  & 1           &1               &    1          &\overline{\Box \raisebox{2.5mm}{}}  &      1      & 1             &    +1  \\
                          \Box  & 1           &1               &\overline{\Box \raisebox{2.5mm}{}}&   1             &      1      & 1             &    -1  \\
                          1     &  \Box       &1               &     1        &      1          &\overline{\Box \raisebox{2.5mm}{}}& 1            &    -1  \\
                          1     &  \Box       &1               &\overline{\Box \raisebox{2.5mm}{}}&      1          &      1      &   1           &    -1  \\
                          1     &  1          &   \Box         &      1       &      1          &\overline{\Box \raisebox{2.5mm}{}}&    1         &    +1  \\
                          1     &  1          &   \Box         &      1       & \overline{\Box \raisebox{2.5mm}{}}  &      1       &    1         &    +1  \\
\end{array}
\end{equation}
while the spectrum of heavy antiquarks is
\begin{equation} 
  \label{eq:pqstar-masses-antiq}
\begin{array}{ccc|cccc|c}
                     U_{F_L-N_c}  & U_{f-F_L}   &   U_{F_R-f} &   SU_{F_R-f}^L &   SU_{f-F_L}^L &   SU_{F_L}^L &   SU_{F_L}^R & sgn(\mu)\\
\hline
\tilde                    1      &\overline{\Box \raisebox{2.5mm}{}}&  1          &    1          &     1           &      1      & \Box          &    +1  \\
\tilde                    1      & 1           &\overline{\Box \raisebox{2.5mm}{}}&    1          &     1           &      1      & \Box          &    -1  \\
\end{array}
\end{equation}
and the heavy components of the mesons are given by
\begin{equation} 
  \label{eq:pqstar-masses-M}
\begin{array}{ccc|cccc|c}
                     U_{F_L-N_c}  & U_{f-F_L}   &   U_{F_R-f} &   SU_{F_R-f}^L &   SU_{f-F_L}^L &   SU_{F_L}^L &   SU_{F_L}^R & sgn(\mu)\\
\hline
                          1     &  1          &  1             &      1       &          \Box   &      1       &\overline{\Box \raisebox{2.5mm}{}}&    -1  \\
                          1     &  1          &  1             &      \Box    &        1        &      1       &\overline{\Box \raisebox{2.5mm}{}}&    +1  \\
\end{array}
\end{equation}
Here we use the short-hand notation $U_N^{L,R} \equiv U(N)_{L,R}$.
The effective CS couplings are $0$, $- \frac12 (f-F_L)$ and $\frac12 (F_R-f)$ for the three surviving gauge groups, respectively.
Note that the last two sectors are chiral-like theories of the type 
$U(k)_{\frac{k}{2}}$ and $U(k)_{-\frac{k}{2}}$ with $k$ antifundamentals as the example discussed in section \ref{sec:simple}.
They are hence both dual to a singlet with topological charge $\pm 1$, 
which can be interpreted as the electric monopoles \cite{Intriligator:2013lca}.

Knowing the spectrum of heavy fermions $\eqref{eq:pqstar-masses-q}-\eqref{eq:pqstar-masses-M}$ 
we can calculate the CS terms as in the last sections.
Together with the extra phase that we get from dualizing the two chiral sectors into monopoles,
one obtains the relative contact terms of the $(p,q)^*$-duality \cite{Benini:2011mf}.

\section{Dualities with tensor matter}
\label{newtoto}
In this section we consider gauge theories with anti(fundamental) and
adjoint matter. We consider cases with $U(N_c)$ and $SP(2N_c)$ gauge
groups, with and without CS terms.  We choose the interactions, the
field content and the values of the CS levels such that the models are
superconformal field theories (SCFTs) and the dual phases have a
vanishing gauge group.  These dualities are generalizations of the one
discovered in \cite{Jafferis:2011ns} and further studied in
\cite{oai:arXiv.org:1110.2547,Agarwal:2012wd}.

The dualities that we introduce can be connected by an RG flow.  We
will show that the dual partition functions are identical, due to some
mathematical identities summarized in the appendix.  
Moreover we show that the contact terms reproduce the complex phase in these identities.

\subsection{Dualities among $U(N_c)$ models}
In this section we study a set of dualities between $U(N_c)$ theories
and theories of singlets, with a vanishing gauge group.

\subsubsection*{$U(N_c)_0$ with one fundamental, one anti fundamental
  and an adjoint field}

Here we consider a $U(N_c)$ gauge theory with a fundamental $Q$, an
antifundamental $\widetilde Q$ and an adjoint field $X$. The
superpotential and the CS level are vanishing.  The fundamentals have
charge $1$ under $U(1)_A$ and $R$-charge $\Delta$.  The adjoint has
$R$-charge $\Delta_X$.

This theory is dual to a set of singlets, with vanishing gauge group.
We identify three types of singlets.
\begin{enumerate}
\item The mesons of the electric theory.  
  They are gauge invariant combinations of the matter fields of the form
\begin{equation}
M_j = Q X^{j} \widetilde Q \quad\quad \text{with }j=0,\dots,N_c-1
\end{equation}
\item A set of operators $U_j=$Tr$X^{j+1}$, for $j=0,\dots,N_c-1$
\item The ``dressed'' monopole operators of the electric theory.  They
  have been first analyzed in \cite{Kim:2013cma} in a similar context.
  Again there is a whole tower of singlets: the usual ``bare'' electric
  monopole operators of Aharony duality $T_{0}$ and $\widetilde T_0$
  are dressed with additional matter fields $T_{j} = \tr(T_{0}X^j)$
  and $\widetilde T_{j} = \tr(\widetilde T_{0}X^j)$ with
  $j=0,\dots,N_c-1$.
\end{enumerate}
The singlets   and their charges under the global symmetries are
\begin{equation} 
\begin{array}{c||ccc}
               & U(1)_A & U(1)_R & U(1)_j \\
\hline
U_j&0&(j+1)\Delta_X&0\\
M_j&2&2 \Delta+j \Delta_X&0\\
T_{j}&-1 &(1-\Delta)-\Delta_X(N_c-j-1) & 1  \\
\widetilde T_{j}&-1 &(1-\Delta)-\Delta_X(N_c-j-1) &- 1  \\
\end{array}
\end{equation}
As a first check, we use the mathematical identity $\eqref{eq:IntId-unit-Aha-Adj}$ to 
show that the two partition functions of the electric and the magnetic theory are identical.
Similar to Aharony duality studied in section \ref{sec:contact-unitary} this matching does not involve any complex phase.
We verify the absence of the phase by computing the relative contact terms as above.
Integrating out all matter fields we have
\begin{equation} \label{eq:contact-unitary-Aha-Adj}
  \begin{aligned}
  & k_{rr}^m = \frac12 \sum_{j=0}^{N_c-1} ( (\Delta_X j + 2 \Delta -1)^2 - 2(\Delta+\Delta_X(N_c-1-j))^2+(\Delta_X (j+1)-1)^2) \\
  & k_{rr}^e = N_c (\Delta-1)^2 + \frac12 N_c^2 (\Delta_X -1)^2  \equiv k_{rr}^m + \frac12 N_c^2 \\
  & k_{AA}^m = \frac12 \sum_{j=0}^{N_c-1} ( 4 - 2 ) \equiv k_{AA}^e =
  N_c \hspace{2cm} k_{JJ}^m = -\frac12 \sum_{j=0}^{N_c-1} ((+1)^2 +
  (-1)^2) = -N_c 
  \end{aligned}
\end{equation}
and we are left with a pure CS gauge theory with effective level $1$.
Adding, as in section \ref{sec:simple}, the contact terms of this theory we find that the overall relative contact terms vanish.
Note that the effective CS level for the background vector
of $U(1)_J$ is $k_{JJ}^e = -N_c$ \cite{oai:arXiv.org:1206.5218}.

\subsubsection*{$U(N_c)_{1/2}$ with one fundamental  and an adjoint field}

In this section we obtain a second dual pair by flowing from the
duality discussed in the last section.  The flow is generated by
assigning a positive real mass to the antifundamental $\widetilde Q$.
We remain with an $U(N_c)$ gauge theory with one fundamental $Q$, one
adjoint $X$ and CS level $k=1/2$, generated at one loop.  In the dual
theory the mesons $M_j$ and the antimonopoles $\widetilde T_j$ acquire a mass.  The
dual theory is a collection of singlets $U_j$ and monopoles $T_{j}$.
The global charges of the singlets in this case are
\begin{equation} 
\begin{array}{c||ccc}
  & U(1)_A & U(1)_R & U(1)_j \\
  \hline
  U_j&0&(j+1)\Delta_X&0\\
  T_{j}&-\frac12  & \frac12 (1-\Delta)-\Delta_X(N_c-j-1) &
 1 \\
\end{array}
\end{equation}
We  check this duality by matching the electric and the magnetic partition functions.
Indeed, this reproduces the mathematical integral identity $\eqref{eq:lafigapuzzadiostriche}$,
up to a complex phase.
This phase can again be obtained from the contact terms, the computation is very similar as in 
the warm-up example of section \ref{sec:simple}.
Due to the chiral matter content the topological symmetry of the pure CS gauge theory 
is shifted
\begin{equation} 
  U(1)_{\widetilde J} = U(1)_J + \frac12 U(1)_A + \frac12 (\Delta-1) U(1)_r
\end{equation}
The final result for the relative contact terms is 
\begin{equation} \label{eq:contact-unit-chiral-adj} 
  \begin{aligned}
\Delta k_{rr} &=  \frac18 N_c (1 - \Delta) (3 \Delta +1 + 2 (N_c-1) \Delta_X) 
&\Delta k_{rJ} &=  \frac{1}{8} N_c (3 \Delta -1 +(N_c-1) \Delta_X )\\
\Delta k_{AA} &= -\frac38 N_c
&\Delta k_{AJ} &= \frac38 N_c\\ 
\Delta k_{rA} &= \frac{1}{8} N_c (1-3 \Delta +(1-N_c) \Delta_X ) 
&\Delta k_{JJ} &= \frac18 N_c
  \end{aligned}
\end{equation}
Using $\eqref{eq:phase}$ these contact terms reproduce indeed the complex phase of $\eqref{eq:lafigapuzzadiostriche}$.

\subsubsection*{$U(N_c)_1$ with an adjoint field}

By integrating out the remaining fundamental, again with positive mass, we obtain 
a $U(N_c)_1$ gauge theory with an adjoint $X$.
The magnetic dual reduces to the set of singlets $U_j$.
This duality has already been studied by \cite{oai:arXiv.org:1110.2547},
generalizing the one discovered in \cite{Jafferis:2011ns}.

In \cite{oai:arXiv.org:1110.2547} the duality has been checked by matching the superconformal indices of the two phases.
Moreover the matching of the partition functions has been performed in \cite{Agarwal:2012wd} based on 
the integral identity $\eqref{eq:eperquestoiononlamangio}$.

The two dual partition functions are again identified up to a complex phase. 
Here we verify that this phase can be deduced from the fermionic spectrum.
The relative contact terms are
\begin{equation}
\begin{array}{ll}
  \Delta k_{rr} = N_c \left(\frac{1}{3} \Delta_X ^2 \left(N_c-1\right) \left(2 N_c-1\right)+2 \Delta_X  \left(N_c-1\right)+2\right)
  \quad \quad \quad&
  \Delta k_{JJ} = -N_c
\end{array}
\end{equation}
and they indeed give rise to the complex phase in $\eqref{eq:eperquestoiononlamangio}$ via $\eqref{eq:phase}$.\\
\\

Let us make two additional comments. 
We can connect these dualities to other dualities studied in
the literature.  As observed in \cite{oai:arXiv.org:1110.2547} by
adding a superpotential deformation $W=Tr X^{N_c+1}$ one can recover
the limiting case of the duality of
\cite{oai:arXiv.org:0808.2771,oai:arXiv.org:0903.0435} in the case
with $k=1$ and $F=0$, where indeed $\widetilde N_c=0$. Also the first
duality that we discussed above can be deformed in the same way. In
this case one recovers the limiting case of the duality studied in
\cite{Kim:2013cma}, still with $\widetilde N_c=0$.

We assigned a generic value $\Delta_X$ to the $R$-charge of the adjoint
$X$. It follows that the singlets $U_j$ have $R$-charge $j \Delta_X$,
even if free chiral multiplets have usually $R$-charge $1/2$ in three
dimensional SCFTs.  This signals that the exact superconformal $R$-symmetry
is accidental and one should compute it as in \cite{Agarwal:2012wd}. In our
calculation we do not discuss the case with the exact $R$-charges and
assign the UV value to the singlets $U_j$.

\subsection{Dualities among $SP(2N_c)$ models}

Dualities with tensor matter exist also for symplectic gauge groups
\footnote{Here we consider the convention $SP(2)_{2k} = SU(2)_k$.} .
We consider $SP(2N_c)_{2k}$ gauge theories with $2F$
fundamental quarks $Q$ and a totally antisymmetric tensor $X$.
We study the following classes of theories: $(2k,2F) =
\{(0,4),(1,3),(2,2),(3,1),(4,0) \}$. The last case has been already
introduced in \cite{oai:arXiv.org:1110.2547}.  Observe that as before
one can start from the case with $2F=4$ and $2k=0$ and obtain the
others by integrating out matter fields.

All the dual phases have a vanishing gauge group and the field 
content consists of a set of singlets. In the first case $k=0$ there are mesons 
$M_i^{ab} \sim Q^a J (XJ)^iQ^b$, monopoles $T_j$
and singlets $U_j=\tr (XJ)^{j+1}$,  with
$J$ being the invariant tensor of the symplectic group.
In the cases with $2k=1$ and $2k=2$ there are no monopoles in the dual
phase, but only mesons and singlets $U_i$.
In the cases with $2k=3$ and $2k=4$ there are just the singlets $U_j$.
\begin{equation}
\begin{array}{c||ccc}
&SU(2F) & U(1)_A &U(1)_r \\
\hline
\hline
Q&2F&1&\Delta \\
X&1&0&\Delta_X\\
\hline
M_j&F(2F-1)&2&\Delta_X+2 \Delta\\
T_j&1&-F&2-4\Delta-\Delta_X(N_c-1+j)\\
U_j&1&0&(j+1) \Delta_X\\
\end{array}
\end{equation}

As before we can test these dualities with the partition function. In
appendix \ref{IdU(N)freeduals2} we review the mathematical identities
discovered in \cite{VdB}.  
We observe that in all the models listed above the partition functions match up to complex phases.
We computed also the relative contact terms in each duality and found that they reproduce these complex
phases through $\eqref{eq:phase}$.

In the case with $2k =0$ and $2 F = 4$ the relative contact terms
vanish and the phase is absent. 
In the other cases we can write the relative abelian contact terms as
\begin{eqnarray}
\Delta k_{rr} &=& 
\frac{1}{2} \left(2 F-1\right) F \sum _{j=0}^{N_c-1} \left(2 \Delta +j \Delta _X-1\right)^2+\frac{1}{2} \sum _{j=0}^{N_c-1} \left((j+1) \Delta _X-1\right)^2
\nonumber \\
&-&2 (\Delta -1)^2 N_c F-\frac{1}{2} N_c \left(2 N_c-1\right) \left(\Delta _X-1\right){}^2+\frac{1}{2} \left(2 N_c+1\right) N_c \nonumber \\
\Delta k_{rA}&=&
\frac{1}{2} N_c F (-4 \Delta +(2 F-1) (4 \Delta +(N_c-1) \Delta_X -2)+4) \nonumber \\
\Delta k_{AA}&=& 4 N_c F(F-1)
\end{eqnarray}
For $2F=2,3$ we also have a non-abelian flavour symmetry with relative contact terms
$\Delta k_{SU(2F)} = N_c (F-2)$.
\section{Dualities between $U(N)$ and $SP(2N)$ theories}
\label{Sec:newduality}

In this section we propose some new dualities between gauge theories
with unitary and symplectic gauge groups. These dualities are
supported by some integral identities between their three sphere
partition functions that we review in appendix \ref{appUSP}.  We
perform another check of these dualities by matching the relative
contact terms with the complex phases.

\subsubsection*{$U(N_c)_0$ and $SP(2N_c)_2$}
\begin{itemize}
\item The electric theory is an $U(N_c)_0$ gauge theory with two pairs of
quarks and  antiquarks and with an
adjoint field $X$.
\item The magnetic theory is an $SP(2N_c)_{2}$ gauge theory with four
flavors and a totally
antisymmetric tensor $Y$.  There are extra
singlets $M_j$ and $N_j$  interacting through the superpotential 
\begin{equation}
\label{symbreW}
\Delta W = 
\sum_{j=0}^{N_c-1}
\left( 
M_j q J (Y J)^{N_c-1-j} q
+
N_j p J ( YJ)^{N_c-1-j} p  
\right)
\end{equation}
\end{itemize}
The electric theory has a global
$SU(2)_L \times SU(2)_R \times U(1)_A \times U(1)_r \times U(1)_J$
symmetry.  The matter fields have charges 
\begin{equation} \label{eq:charges-USPe}
\begin{array}{c||ccccc}
& SU(2)_L & SU(2)_R & U(1)_A & U(1)_r & U(1)_J \\
\hline
Q &2&1&1&\Delta_Q&0\\
\widetilde Q &1& 2&1&\Delta_Q&0\\
X&1&1&0&\Delta_X&0
\end{array}
\end{equation}
The global flavor symmetry of the dual phase is $SU(4)$.  This
symmetry is broken to $SU(2)^2 \times U(1)_V$ by the superpotential
(\ref{symbreW}). There are also an axial $U(1)_A$ and an $U(1)_r$
$R$-symmetry.  The charges of the matter fields are
\begin{equation} \label{eq:charges-USPm}
\begin{array}{c||ccccc}
& SU(2)_L & SU(2)_R & U(1)_A & U(1)_r & U(1)_V \\
\hline
q &2&1&1&\Delta_q&-1\\
p &1& 2&1&\Delta_p&\,\,\,\,\,1\\
Y&1&1&0&\Delta_Y&\,\,\,\,\, 0\\
M_j&1&1&-2&2(1-\Delta_q)-\Delta_Y(N_c-1-j)&\,\,\,\,\, 2 \\
N_{j}&1&1&-2&2(1-\Delta_p)-\Delta_Y(N_c-1-j)&\,\,- 2 \\
\end{array}
\end{equation}
where the global charges of the singlets $M_j$ and $N_j$ have been computed from (\ref{symbreW}).
These charges coincide with the ones of the electric monopoles $T_i$ and $\widetilde
T_i$ after the identifications 
\begin{align} 
  \Delta_Q = \Delta_q = \Delta_p = \Delta \hspace{1.5cm} \Delta_Y = \Delta_X
\end{align}
and 
\begin{align}
   U(1)_J = 2 \, U(1)_V 
\end{align}
such that $\lambda=4 \phi$.
Note that the topological symmetry of the unitary theory appears as a flavour symmetry in the symplectic side.

By inserting the real masses according to the global charges 
in the relation (\ref{USP10}) we conclude that the two models
have the same partition function modulo an extra complex phase.

In the rest of this section we compute the CS contact terms for the
global symmetries and match the result with the phase in
(\ref{USP10}).  In the unitary case we have
\begin{equation}
\begin{aligned}
k_{rr}^e     &=& -\frac12 N_c^2+2N_c(\Delta-1)^2
+\frac12 N_c^2(\Delta_X-1)^2 ,\quad
k_{AA}^e     =  2 N_c,\quad
k_{JJ}^{e} = -\frac12 N_c \\
k_{rA}^e     &=&  2N_c(\Delta-1),\quad
k_{rJ}^e     =  0 ,\quad
k_{AJ}^e     =  0 ,\quad
k_{SU(2)_L}^e =  \frac12 N_c,\quad
k_{SU(2)_R}^e =  \frac12 N_c  
\end{aligned}
\end{equation}
while in the dual symplectic case we have
\begin{equation}
\begin{aligned}
&k_{rr}^m     =   -\frac14 N_c(N_c+1)+2 N_c(\Delta-1)^2 
+\frac12 N_c(N_c-1)(\Delta_X-1)^2+\sum_{j=0}^{N_c-1}(1-2\Delta-j\Delta_X)^2
\\
&k_{rA}^m     =  N_c(1-\Delta)+
2\sum_{j=0}^{N_c-1}(1-j\Delta_X-2\Delta),\quad
k_{VV}^{m} = 0\\
&k_{AA}^m     =-2N_c ,\quad 
k_{rV}^m     = 0 ,\quad
k_{AV}^m =  0,\quad
k_{SU(2)_L}^m = N_c,\quad
k_{SU(2)_R}^m =  N_c
\end{aligned}
\end{equation}
These contact terms  reproduce the phase in (\ref{USP10}).

\subsubsection*{$U(N_c)_1$ and $SP(2N_c)_4$}

We can obtain another duality from this by integrating out some matter fields.
In the unitary phase we integrate out one quark and one antiquark, with positive real mass. 
In the dual phase we integrate out one $q$ quark, one $p$ quark and the monopoles.
We end up with the following two theories.
\begin{itemize}
\item The electric theory is an $U(N_c)_1$ gauge theory with one
quark and one anti-quark and an
adjoint field $X$.
\item The magnetic theory is an $SP(2 N_c)_{4}$ gauge theory with two
flavors and with a totally antisymmetric
tensor $Y$.
\end{itemize}
The global symmetry of the unitary phase is $ U(1)_A\times U(1)_r \times U(1)_J $
with charges as in table $\eqref{eq:charges-USPe}$.
In the symplectic dual we have $U(1)_A \times U(1)_r \times U(1)_V$ symmetry
with charges $\eqref{eq:charges-USPm}$.
We again identify the global $U(1)_V$ of the magnetic phase with the topological $U(1)_J$ of the electric phase as 
$U(1)_J = 2 \, U(1)_V$ such that $4 \phi= \lambda$.
The partition functions of the two models are related by the integral identity (\ref{USP21}),
up to the usual complex phase.

To match also the phase we integrate out the matter fields and flow to a pure CS theory.
The contributions to the contact terms from the electric phase are
\begin{equation}
\begin{aligned}
\label{ele21}
k_{rr}^{e} &=& -\frac12 N_c^2 + N_c(\Delta-1)^2+\frac12 N_c^2(\Delta_X-1)^2
,\quad
k_{AA}^{e} = N_c \\
k_{JJ}^{e} &=& -\frac12 N_c,\quad
k_{rA}^{e} = N_c(\Delta-1),\quad
k_{rJ}^{e} = 0,\quad
k_{AJ}^{e} = 0
\end{aligned}
\end{equation}
With this choice the contact terms
of the magnetic phase are
\begin{equation}
\begin{aligned}
\label{magn21}
k_{rr}^{m} &=& -\frac12 N_c(2N_c+1)+ 2 N_c (\Delta-1)^2
+N_c(2N_c-1)(\Delta_X-1)^2 \\
k_{AA}^{m} &=& 2N_c,\quad
k_{VV}^{m} = N_c,\quad
k_{rA}^{m} =N_c(\Delta-1)\quad
k_{rV}^{m} =0\quad
k_{AV}^{m} =0
\end{aligned}
\end{equation}
The relative contact terms obtained from (\ref{ele21}) and
(\ref{magn21}) reproduce the complex phase in (\ref{USP21}).

\section{Conclusions}
\label{Sec:conclusions}

In this paper we computed the relative contact terms of two point
functions of global symmetries for pairs of dual three dimensional
$\mathcal{N}=2$ theories.  The calculation provides an extra check of
these dualities.  We found agreement with the results obtained from
localization.
We also generalized some known dualities and proposed a new one, between symplectic and unitary gauge groups.

Extensions of our results are possible. First it would be
interesting to study the dualities with tensor matter and a polynomial
superpotential
\cite{oai:arXiv.org:0808.2771,oai:arXiv.org:0903.0435,Morita:2011cs,oai:arXiv.org:1110.2547,Kim:2013cma}.
In those cases the integral identities between the dual partition
functions have, to our knowledge, not been studied in the literature.
By computing the contact terms one can
determine the possible complex phase in these identities and hence predict
some new mathematical relations as the ones derived in \cite{VdB}.  An
independent derivation would be given by dimensional reduction from
the four dimensional superconformal index
\cite{Romelsberger:2005eg,Kinney:2005ej} along the lines of
\cite{Aharony:2013dha}.

The duality between unitary and symplectic gauge theories requires further investigation.
First, it would be desirable to formulate it for arbitrary values of $N_c$, $N_f$ and $k$.
Then, additional checks are necessary, such as computing the superconformal
index and the Witten index after having lifted the moduli space.
Other dualities between unitary and real groups have already been
suggested in the literature \cite{Aharony:2008gk}.  It would be
interesting to connect this scenario to our results.

We would like to conclude with an observation on the Aharony like
dualities, i.e. dualities with vanishing CS levels.
In all these cases there is no extra phase in the integral identities or, equivalently,
the relative contact terms vanish.
One may wonder if this is related to the parity invariance of the theory.\footnote{
We are grateful to Ken Intriligator for interesting comments on this point.}

\section*{Acknowledgements}

It is a great pleasure to thank Ken Intriligator and Alberto Zaffaroni for useful discussions and 
interesting comments on the draft.
A.A. is supported by the Institut de Physique Th\'eorique Philippe Meyer at the
\'Ecole Normale Sup\'erieure.
The work of C.K. is supported by ANR grant 12-BS05-003-01.

\appendix

\section{Partition function and contact terms}
\label{sec:Part-func}

In this appendix we report some results concerning the localized
partition function on the squashed three sphere computed in
\cite{oai:arXiv.org:1102.4716} from localization\footnote{ See also
  \cite{Kapustin:2009kz,Jafferis:2010un,Hama:2010av} for related
  computations on the round sphere.}.  We first review the definition
of the matrix integral. Then we report some integral identities
between pairs of matrix integrals.  They represent the relation
between the dual pairs studied in the paper.  We conclude by
discussing the extra complex phases appearing in these identities
and their relation with the contact terms.

\subsection{The partition function on the squashed three sphere}
The partition function of a three dimensional $\mathcal{N}=2$
supersymmetric CS matter theory localized on a squashed three
sphere $S_b^3$ corresponds to a matrix integral. For a general gauge group $G$ 
it can be written as
\begin{eqnarray}
\label{eq:Zdef}
Z_{S_b^3} = 
&&\frac{1}{|W|} \int \prod_{i=1}^{\text{rank} [G]} d \sigma_i \, \,
Z_V \,\,\,
\prod_{I \in \mathcal{R}} Z_I 
\, \,\,
c\left(2 \lambda Tr \sigma - 2 k Tr \sigma^2 \right) 
\end{eqnarray}
Let us briefly review the different factors.  The integral is over the
Cartan of the gauge group, here parameterized by the diagonal entries
$\sigma_i$ of the real scalar $\sigma$ in the vector multiplet of the gauge
field.  The function $c(x)$ is
\begin{equation}
  \label{eq:c-of-x}
c(x) = e^{\frac{i \pi x}{2\omega_1 \omega_2}}
\end{equation}
where $\omega_1 = i b$ and $\omega_2 = i / b$ with $b$ being the
squashing parameter of the sphere $S_b$.  The term $c\left(2 \lambda
  Tr \sigma - 2 k Tr \sigma^2 \right)$ is the contribution from the classical
action, a CS term at level $k$ and an FI term
with parameter $\lambda/2$.  $|W|$ represents the sum over the Weyl
degeneracies.  The one loop determinants $ Z_V$ and
$Z_I$ are the contributions from the vector and the
matter multiplet respectively,
\begin{eqnarray} 
  \label{eq:Zs-def} 
Z_I= \prod_{\rho_I, \tilde \rho_I} \Gamma_h\left(
\omega \Delta_I + \rho_I(\sigma)+\widetilde \rho_I(\mu)\right)
,\quad \quad
Z_V
=
\prod_{\alpha}\Gamma_h^{-1}\left(\pm \alpha(\sigma)\right)
\end{eqnarray}
where 
$\Gamma_h(\pm x) \equiv \Gamma_h(x)\Gamma_h(-x)$.
The ``hyperbolic Gamma function'' $\Gamma_h$, which features
prominently in these expressions, has been the interest of recent
mathematical research, see \emph{e.g.\ }\cite{VdB} for which we refer
for further details.  It can be written as
\begin{equation}
\label{eq:Gammahvbd}
\Gamma_ h(z;\omega_1,\omega_2) \equiv \Gamma_h(z) \equiv
\prod_{n,m=1}^{\infty}
\frac{(n+1)\omega_1+(m+1) \omega_2-z }{n \omega_1+m \omega_2+z}
\end{equation}
Note that in most expressions we suppress the periodicities $\omega_1$
and $\omega_2$ which for us will always be as in $\eqref{eq:c-of-x}$.
The contribution of the vector multiplet is parameterized by the
positive roots $\alpha$ of the gauge group.  The matter sector gets a
contribution from each chiral multiplet labeled by $I$ in the 
representation $\mathcal{R}$.  It depends on
the weights $\rho_I$ of the gauge representation, the
weights $\tilde \rho_I$ of the flavor representation and the $R$
charge $\Delta_I$.  We also defined $\omega \equiv (\omega_1 +
\omega_2)/2$.

\subsection{Integral identities}

\subsubsection*{$U(N_c)$ models with an adjoint}
\label{IdU(N)freeduals}
In this appendix we report the mathematical identities discovered in
\cite{VdB} that show the matching between the partition function of
the $U(N_c)$ gauge theories with adjoint matter and dual theories made
of free singlets. We discussed these dualities in section \ref{newtoto}.

The first identity relates a $U(N_c)_0$ gauge theory with one
fundamental, one antifundamental and an adjoint to a set of mesons,
dressed (anti-)monopoles and singlets $U_j$. We
have
\begin{eqnarray} \label{eq:IntId-unit-Aha-Adj}
Z_{U(N_c)_0}(\mu,\nu,\tau,\lambda)&=&
\prod_{j=0}^{N_c-1} 
\Big(
\Gamma_h\big(\omega\pm
\frac12 \lambda-\frac12 (\mu+\nu)- \omega \tau(N_c-1-j) 
\big)
\nonumber \\
&\times &
\Gamma_h((j+1) \tau)\quad
\Gamma_h(j \omega \tau+\mu+\nu)\phantom{\frac{1}{2}}\!\!\!\!
\Big)
\end{eqnarray}
where $\mu=\nu=m_A + \omega \Delta$.

The second identity relates a $U(N_c)_{1/2}$ gauge theory with 
one fundamental and an adjoint to a set
of dressed monopoles and singlets $U_j$.
\begin{eqnarray}
  \label{eq:lafigapuzzadiostriche}
Z_{U(N_c)_{\frac{1}{2}}}(\mu,\tau,\!\!&&\lambda)=
\prod_{j=0}^{N_c-1} \Big(\!\Gamma_h((j\!+\!1) \tau)
\; \;
\Gamma_h\big(
\frac12 (\omega+\lambda-\mu)\!-\!  \omega\tau (N_c\!-\!1\!-\!j)
\big)
\Big)
\\
\times&&  
\zeta^{-Nc}    
c\Big(
\frac14 N_c\big(
(\lambda \!-\! \omega)^2 \!-\!3 \mu^2 \!+\!2\mu(3 \lambda\!+\!\omega)
+2 \omega \tau (N_c-1)(\lambda+\omega-\mu)
\big)
\!\Big)
\nonumber
\end{eqnarray}
where $\mu=m_A + \omega \Delta$.

The last identity relates a $U(N_c)_{1}$ gauge theory with 
an adjoint to a set of singlets $U_j$.
\begin{eqnarray}
  \label{eq:eperquestoiononlamangio}
Z_{U(N_c)_1}(\tau,\lambda)&=& \zeta^{-3 Nc}
\prod_{j=0}^{N_c-1} \Gamma_h((j+1) \tau) \\
&\times& 
c\Big(\frac12 N_c \big(
2 \omega^2 +\omega^2 \tau +\lambda^2+2(N_c-1) 
+\frac13 \omega^2 \tau^2
(N_c-1)(2N_c-1)
\big)\Big)
\nonumber
\end{eqnarray}

\subsubsection*{$SP(2N_c)$ models with a totally antisymmetric field}
\label{IdU(N)freeduals2}
In this appendix we report the mathematical identities \footnote{In
  some case we manipulate the identities such that they appear more
  directly related to the physical dualities that we discussed in the
  paper.}, discovered in \cite{VdB}, that show the matching between
the partition function of the $SP(2N_c)$ gauge theories with totally
antisymmetric matter and dual theories made of free singlets. We
discussed these dualities in section \ref{newtoto}.
The first identity relates a $SP(2N_c)_0$ gauge theory with four
fundamentals and a totally antisymmetric field to a set of mesons,
dressed monopoles and singlets $U_j$. 
\begin{eqnarray}
Z_{SP(2N_c)_0}(\mu,\tau,\lambda)  
&=&
\prod_{j=0}^{N_c-1} 
\Big(
\Gamma_h((j+1) \tau)
\Gamma_h\big(2 \omega -\sum_{\alpha=1}^{4} \mu_\alpha
-\omega \tau (N_c+j-1)
\big)
\nonumber \\
&\times &
\prod_{0 \leq \alpha < \beta \leq 3}
\Gamma_h(j \omega \tau+\mu_\alpha+\mu_\beta)\phantom{\frac{1}{2}}\!\!\!\!
\Big)
\end{eqnarray}
where the four components of the vector $\mu$ are all equal to
$m_a+\omega \Delta$.

The second identity relates a $SP(2N_c)_1$ gauge theory with  three
fundamentals and a totally antisymmetric field to a set of mesons and
singlets $U_j$. We have
\begin{eqnarray}
&&Z_{SP(2N_c)_1}(\mu,\tau,\lambda) 
=
\prod_{j=0}^{N_c-1} 
\Big(
\Gamma_h((j+1) \tau)
\prod_{0 \leq \alpha < \beta \leq 2}
\Gamma_h(j \omega \tau+\mu_\alpha+\mu_\beta)
\Big)
 \\
\times &&
c\Big(N_c \big(2 \prod_{0 \leq \alpha < \beta \leq 2} \mu_\alpha \mu_\beta
+2 \omega \tau (N_c-1)\sum_{\alpha=1}^{3} \mu_\alpha+
\frac13 \omega^2 \tau^2(N_c-1)(4N_c-5)
\big)
\Big)\nonumber
\end{eqnarray}
where the three components of the vector $\mu$ are $m_A+\omega
\Delta$.

The third identity relates a $SP(2N_c)_{2}$ gauge theory with two
fundamentals and a totally antisymmetric field to a set of mesons and
singlets $U_j$. We have
\begin{eqnarray}
&&Z_{SP(2N_c)_2}(\mu,\tau,\lambda) 
=
\prod_{j=0}^{N_c-1} 
\Big(
\Gamma_h((j+1) \tau)
\Gamma_h(j \omega \tau+\mu_1+\mu_2)
\Big)
 \\
\times &&
c\Big(N_c \big(2 \omega(\mu_1+\mu_2)+\omega \tau (N_c-1)(\mu_1+\mu_2+2 \omega)
+\frac23 \omega^2 \tau^2(N_c-1)(N_c-2)
\big)\Big)\nonumber
\end{eqnarray}
The fourth identity relates a $SP(2N_c)_{3}$ gauge theory with one
fundamentals and a totally antisymmetric field to a set of
singlets $U_j$. We have
\begin{eqnarray}
&&Z_{SP(2N_c)_3}(\mu,\tau,\lambda) 
=
\zeta^{- N_c}\prod_{j=0}^{N_c-1} 
\Gamma_h((j+1) \tau)
 \\
\times &&
c\Big(N_c \big(4 \mu \omega-2 \mu^2+\omega^2+3\omega^2 \tau (N_c-1)
+\frac16 \omega^2 \tau^2(N_c-1)(2N_c-7)
\big)
\Big)\nonumber
\end{eqnarray}
where $\mu=m_A+\omega \Delta$.

The last identity relates a $SP(2N_c)_{4}$ gauge theory with
 a totally antisymmetric field to a set of
singlets $U_j$. We have
\begin{eqnarray}
&&Z_{SP(2N_c)_4}(\tau,\lambda) 
=
\zeta^{-3N_c}\prod_{j=0}^{N_c-1} 
\Gamma_h((j+1) \tau)
 \\
\times &&
c\Big(N_c \big(3 \omega^2(\tau(N_c-1)+1)
+\frac 16 \omega^2 \tau^2(N_c-1)(2N_c-7)
\big)
\Big)\nonumber
\end{eqnarray}

\subsubsection*{U/SP relations}
\label{appUSP}

\subsubsection*{$U(N_c)_0$ and $SP(2N_c)_1$}
\begin{eqnarray}
\label{USP10}
&&
Z_{U(N_c)_0}(\mu;\nu;\tau;\lambda)
\!=\!
Z_{SP(2N_c)_1}(\mu_{\sigma'};\tau)
\!
\prod_{j=0}^{N_c-1}
\!
\Gamma_h
\Big(\!
2\omega\pm \frac12 \lambda
\!-\! \frac12 \sum_{\alpha=1,2}
(\mu_\alpha+\nu_\alpha)
\!-\!\tau(N_c\!-\!1\!-\!j)
\!\Big)
\nonumber
\\
&&
c\Big(
N_c\big(
4 {\sigma'}^2
\!-\!2 \mu_1 \mu_2 -2 \nu_1 \nu_2
-(N_c-1)\tau 
\sum_{\alpha=1,2}
\left(\mu_\alpha+\nu_\alpha
\right)
-\frac{2}{3}(N_c-1)(N_c-2)\tau^2
\big)\!\!
\Big)
\nonumber \\
\end{eqnarray}
 where $\mu$ and $\nu$ are $2$-vectors.  Moreover
$4\sigma' = \nu_1+\nu_2-\mu_1-\mu_2-\lambda$ and
$\mu_{\sigma'}=(\mu_1+\sigma',\mu_2+\sigma',\nu_1-\sigma',
\nu_2-\sigma')$. 

\subsubsection*{$U(N_c)_1$ and $SP(2N_c)_2$}
\begin{eqnarray}
\label{USP21}
Z_{U(N_c)_1}(\mu;\nu;&&\!\!\tau;\lambda)
=
Z_{SP(2N_c))_2}
\big(
 (3\mu+\nu-\lambda)/4,
(\mu+3\nu+\lambda)/4;
\tau \big)
\\
\times
&&
c\Big(\frac12 N_c\big(
\lambda^2
+(\mu+\nu)^2
-4(\mu+\nu) \omega
-4(N_c-1)\tau \omega
+2(N_c-1)\tau^2
\big)
\Big)\nonumber 
\end{eqnarray}
 where $\mu$ and $\nu$ are 1-vectors.

\subsection{Localization and contact terms}
Let us summarize here the relationship between the extra phase appearing in most of the mathematical identities displayed in the last section
and the relative contact terms for the corresponding dualities.
As discussed in section \ref{sec:contact} the contact terms are associated with global CS couplings in the Lagrangian.
The saddle point of the background vector multiplet contributing to the localized partition function  
is $A_\mu = \lambda = 0$, $D=i\sigma=\text{const}$. 
We consider generic vevs for the global (non-$R$) symmetry multiplets $\langle \sigma_I \rangle = m_I$.
The $R$ symmetry instead plays a special role. 
Preserving supersymmetry on a curved manifold such as the squashed three sphere requires embedding
$A_\mu^r$ in a gravity multiplet and turning on a particular imaginary background value, 
which is determined by the geometry \cite{Festuccia:2011ws}.

The phase appearing in the mathematical identities that we want to match with our computations 
corresponds to the relative contact terms for two dual theories. 
On the supersymmetric locus the electric and magnetic CS Lagrangians $\eqref{eq:CS-action}$ generate the phase 
\begin{equation}
 \label{eq:phase} 
 c\big(-(2 \omega^2 \Delta k_{rr}
 + 4 \omega \sum_{I} \Delta k_{rI}
 + 2 \sum_{I,J} \, \Delta k_{IJ} \, m_I m_J) \big)
\end{equation} 
see also \cite{Benini:2011mf} for a related discussion.
Note that $\omega= \frac i2 (b + \frac1b)$ is related to the background value of the $R$ symmetry
and that $\Delta$ denotes the difference between the two dual pairs, $\Delta k =  k^m - k^e$.
\section{Flowing between dualities}
\label{sec:APPFLOWS}

As discussed in the introduction the dualities of
\cite{oai:arXiv.org:0808.0360} and \cite{Benini:2011mf} can be
obtained from the Aharony duality by a real mass flow. The existence
of reverse flows was not obvious.  A first result in this direction
was obtained in \cite{Intriligator:2013lca} for $k=-1$ and generalized
to generic values of the level in \cite{Amariti:2013qea}. Moreover it
has been shown in \cite{Amariti:2013qea} that one check this flow on
the three sphere partition function.

In this appendix we first review the flow from the 
Giveon-Kutasov to the Aharony duality. 
Then we construct another reverse flow, from the $(p,0)$ to the
Aharony duality and check its validity on the partition function.
As a last example we study a flow connecting two chiral theories,
from the $(p,0)$ duality to the $(p,q)^*$ case.

The charges of the fields under the global symmetries are
\begin{equation} 
\label{eq:charges-table-fundam}
\begin{array}{c||ccccccc}
             &SU(F_L)_L & SU(F_R)_R    & U(1)_A & & U(1)_r && U(1)_J \\
\hline
Q            &   \Box    &   1         &   1 &   & \Delta &&  0     \\
\widetilde Q  &   1    & \overline{\Box \raisebox{2.5mm}{}}&  1 &    & \Delta& &  0     \\
 q            &  \overline{\Box \raisebox{2.5mm}{}}   & 1&-1 & & 1-\Delta &&  0     \\
\widetilde q  &  1    & \Box & - 1 && 1-\Delta &&  0     \\
M&\Box&\overline{\Box \raisebox{2.5mm}{}}&2&&2 \Delta&&0\\
T &1 &1 &-\frac12(F_L + F_R) &\phantom{lt}
&1-N_c+ \frac12(F_L + F_R)(1-\Delta)&\phantom{l}& 1 \\
\widetilde T &1 &1 &-\frac12 (F_L + F_R) 
&&1-N_c+ \frac12 (F_L + F_R)(1-\Delta)&& -1 
\end{array}
\end{equation}
where for vector-like dualities $F_L=F_R=F$.

\subsection{From Giveon Kutasov to Aharony duality }
 
In the electric phase of the Giveon-Kutasov duality the flow is
generated by assigning real masses to the same amount of fundamentals
and antifundamentals. When the number of masses coincides with the CS
level one can choose the opportune sign of the masses to cancel it.
This leaves with the electric phase of the Aharony duality.  The
magnetic side of the flow is more involved. Indeed it is also
necessary to break the gauge symmetry by assigning a non zero vacuum
expectation value to the real scalar $\sigma$ inside the vector
multiplet.  This gauge symmetry breaking gives three gauge
sectors. One of them is the dual gauge sector of the Aharony
duality. The other two sectors can be further dualized to singlets. It
turned out that these singlets have the same quantum numbers of the
electric monopole operators that couple to the magnetic monopole
operator in the dual phase. By considering the gauge sector and these
singlets one reconstruct the Aharony's dual phase.

\subsection{From $(p,q)^*$ to Aharony duality}
\label{app:frompqstartoAharony}
Here we study the flow from the $(p,q)*$ duality to the
Aharony duality.  The electric theory has CS level
$k=s-\frac12 (F_L+F_R)$ (with $F_L>s>F_R$) while the magnetic theory
has CS level $-k$ and $W=M q \tilde q$.  By giving a large real mass
to some of the quarks the $(p,q)^*$ dual pair flows to the Aharony
pair. In the dual theory the gauge symmetry is broken by
shifting the real scalar in the vector multiplet.  In the electric
theory we assign a positive large mass to $F_L-s$ quarks and a
negative one to $s-F_R$. After these fields are integrated out we have
a $U(N_c)_{0}$ gauge theory with $F_R$ light quarks and antiquarks
without superpotential.  This is the 
electric side of the Aharony duality.

In the magnetic theory we assign the masses consistently to the dual
quarks and mesons. Moreover the $\sigma_i$ scalar in the vector
multiplet must get a vev to preserve the duality.  There are $F_R-N_c$
unshifted components. The other $F_L-F_R$ are split. $F_L-s$ acquire a
large negative vev while the remaining $s-F_R$ acquire a large
positive vev.  By integrating out the massive fields one has three
sectors.  The first one has gauge group $U(F_R-N_c)_{0}$ with $F_R$
quarks and antiquarks $q$ and $\tilde q$, a singlet $M$ with $F_R^2$
components and superpotential $W=M q \tilde q$. The second sector has
gauge group $U(F_L-s)_{\frac12 (F_L-s)}$ with $F_L-s$ light chiral
fields and the third one has gauge group $U(s-F_R)_{\frac12(F_R-s)}$
with $s-R_R$ light chiral fields.  As discussed in
\cite{Benini:2011mf,Intriligator:2013lca,Amariti:2013qea} these
sectors are both dual to a single chiral superfield. One couples to
the magnetic monopole the other to the magnetic anti-monopole. They
indeed correspond to the electric monopole and anti-monopole that appear
as singlets in the dual superpotential.

This flow can be studied on the partition function along the lines of
\cite{Amariti:2013qea}. 
The relation between the $(p,q)_a^*$ and the $(p,q)_b^*$ theory is
given in \cite{Benini:2011mf,VdB}.
We turn on the real masses as discussed above and obtain the relation
\begin{eqnarray}
\label{eq:Zpqsab2}
&&
Z_{U(N_c)_{0}}  ( \mu, \nu; \lambda)
=
Z_{U(F_R-N_c)_{0}}
(\omega- \nu,\omega- \mu;- \lambda)
\prod_{\alpha,\beta=1}^{F_R} \Gamma_{h} (\mu_\alpha+\nu_{\beta})
\nonumber \\
&\times & 
c\big(2 \left(\left(m_A-\omega \right) \left(k F_R \left(\omega -3 m_A\right)
-F_L \left(k m_A-3 k \omega +\lambda \right)\right)-
\omega  N_c \left(4 k m_A-4 k \omega +\lambda \right)\right)\big)
\nonumber \\
&\times & 
Z_{U(F_L-s)_{\frac12 (F_L-s)}}
(\mathcal{M}_1,0;\lambda_1)
\quad
Z_{U(s-F_R)_{\frac12 (F_R-s)}}^{(s-F_R,0)}
(\mathcal{M}_2,0;\lambda_2)
\end{eqnarray}
where $\mu$ and $\nu $ are $(F_R)$-vectors. We also have 
an $(F_L-s)$-vector $\mathcal{M}_1$ and an $(s-F_R)$-vector $\mathcal{M}_2$,
with each component equal to $\omega-m_A$.
The shifted FI terms in the two chiral sectors are respectively
\begin{eqnarray}
\lambda_1 &=& - \lambda
-k m_A+\frac12 (F_L m_A)+\frac32 F_R m_A+2 \omega  N_c+2 k \omega 
-F_L \omega -F_L \omega
\nonumber \\
\lambda_2 &=& - \lambda -k m_A-\frac12 F_L m_A-\frac32 F_R m_A
-2 \omega  N_c+2 k \omega +F_L \omega +F_L \omega
\end{eqnarray}
These integral can be computed as in \cite{VdB,Benini:2011mf} and
finally we have
\begin{eqnarray}
Z_{U(N_c)_{0}}  (\mu, \nu; \lambda))
&=&
Z_{U(F_R-N_c)_{0}}
(\omega- \nu,\omega- \mu;- \lambda)
\prod_{\alpha,\beta=1}^{F_R} \Gamma_{h} (\mu_\alpha+\nu_{\beta}) \nonumber
\\
&\times & \Gamma_h \Big(\pm \frac12 \lambda
-m_A F_R +(F_R-N_c+1) \omega\Big)
\end{eqnarray}
the correct relation for the Aharony duality with $N_C$ colors
and $F_R$ flavors.

\subsection{From $(p,q)^*$ to $(p,0)$ duality}
\label{app:frompqstartop0}

In this appendix we consider a second flow from the $(p,q)_*$ duality
to the $(p,0)$ duality.  The electric theory is a $U(N_c)_k$ gauge
theory with $F_R$ light quarks and $F_L$ light antiquarks. The CS
level is $k=s-(F_L+F_R)/2$ and $F_L > s>F_R$.  The dual theory has
gauge group $U(F_L-N_c){-k}$ with $F_R$ antiquarks and $F_L $ quarks,
and $F_L \times F_R$ singlets, the mesons of the electric theory.
For simplicity we fix $k>0$, i.e. $s-F_R>F_L-s$ and integrate out 
$(F_L-s)$ antiquarks with large positive mass. The electric phase
flows to the $(p,0)_b$ theory a $U(N_c)_{(s-F_R)/2}$ CS matter theory
with $F_R$  quarks and $s$ light antiquarks.

In the dual phase one has to assign the masses to the dual quarks and
to the mesons consistently. Moreover one has to assign a vacuum
expectation value to the scalar in the vector multiplet, by giving a
large negative shift to $F_L-s$ components.
In this dual theory we distinguish two sectors. In the first sector we
have gauge sector with $U(s-N_C)_{-(s-F_R)/2}$ gauge symmetry with
$F_R$ light dual antiquarks, $s$ light dual quarks and $s \times F_R$
light singlets, corresponding to the mesons of the electric theory.
The second sector is a $U(s-F_L)_{(s-F_L)/2}$ gauge theory with
$F_L-s$ light quarks. This sector can be dualized to a
monopole. Eventually one obtains the $(p,0)_b$ duality.

This flow can be followed on the partition function.
We start from the equality $(5.19)$ in \cite{Benini:2011mf}
and assign the large masses
and vevs to the fields.
By performing the large mass limit we arrive to the relation
\begin{eqnarray}
\label{eq:Zpq}
&&
Z_{U(N_c)_{\frac12 (s-F_R)}}( \mu, \nu;\lambda)
=
Z_{U(s-N_c)_{-\frac12 (s-F_R)}}(\omega- \nu,\omega- \mu;
-\lambda+(s-F_R) \omega) 
\nonumber \\
&&
\prod_{\alpha=1}^{F_R}\prod_{\beta=1}^{s} \Gamma_{h} (\mu_\alpha+\nu_{\beta})
\quad \quad
Z_{U(F_L-s)_{\frac12 (F_L-s)}} (\mathcal{M},0;\widetilde \lambda)
\nonumber \\
&&
c\Big(
  (s-F_R) \sum _{\beta =1}^s m_{\beta }^2
  +
  \omega  N_c (
  4(F_L-s) m_A +2 \lambda + 5 s \omega -(4 F_L+F_R) \omega) \Big)
\nonumber \\
&&
c\left(\omega -m_A)
  (
  \lambda
  -2 F_L (F_R (m_A-\omega) +2 s \omega)
  +F_L^2 (3 \omega -m_A)
  -s F_R (\omega -3 m_A)
  )
\right)
\nonumber\\
\end{eqnarray}
where $\mu$ is an $(F_R)$-vector and $\nu$ is an $s$-vector.  In this
case there is also an $(F_L-s)$-vector $\mathcal{M}$.  Each component
of $\mathcal{M}$ is equal to $\omega-m_A$.  The shifted FI in the
chiral sector is $\widetilde \lambda=-\lambda + F_L \left(2 \omega
  -m_A\right)-F_R m_A-2 \omega n_c-s \omega +F_R \omega$.  The
partition function of the $U(s-F_L)$ can be explicitly computed. 
We eventually have
\begin{eqnarray}
  \label{eq:integral-p0}
&&Z_{U(N_c)_{\frac12 (s-F_R)}}( \mu, \nu;\lambda)
=
Z_{U(s-N_c)_{-\frac12(s-F_R)}}(\omega- \nu,\omega- \mu;
-\lambda+(s-F_R) \omega) 
\nonumber \\
&&
\prod_{\alpha=1}^{F_R}\prod_{\beta=1}^{s} \Gamma_{h} (\mu_\alpha+\nu_{\beta})
\quad \quad
\Gamma_h 
\Big(-\frac12 \lambda -\frac12 (s+F_R) m_A+\omega \big(\frac12 (s+F_R)
-N_c+1\big)\Big)
\nonumber\\
&&
c\Big(
\left(s-F_R\right) \left(s+F_R\right) \omega  m_A
-s \left(s-F_R\right) m_A^2
-\left(s-F_R\right) \omega ^2 \left(\left(s-N_c\right)\right)
\Big)
\nonumber\\
&&
c\Big(\frac{1}{4} \left(\left(s+F_R\right) m_A
+
2 \omega  N_c-\lambda -3 s \omega +F_R \omega \right)^2
-\left(s-F_R\right) \sum _{\beta =1}^s m_{\beta }^2-\lambda  
\left(s-F_R\right) m_A
\Big)
\nonumber\\
\end{eqnarray}
One can check that this is the right phase of the $(p,0)_b$
duality. It can be obtained from the $(p,0)_a$ by acting with parity
as discussed in \cite{Benini:2011mf}.

\bibliographystyle{JHEP}
\bibliography{BibFile}

\end{document}